# A simple and fast frequency domain analysis method for calculating the frequency response and linearity of electro-optic micro-ring modulators

PAYAM RABIEI

*Partow Technologies LLC, 1487 Poinsettia Ave, Suite 119, Vista, CA 92081*
*pr@partow-tech.com*

**Abstract:** A fast and simple frequency domain method is introduced for the analysis of micro-ring modulator response using the Jacobi–Anger expansion method. Resonance frequency modulated micro-ring (FMMR) modulators and coupling modulated micro-ring modulators (CMMR) are analyzed using this method. The linearity of these modulators is analyzed. The third order intercept point (IP$_3$) is calculated for CMMR devices and compared to Mach-Zehnder interferometer (MZI) modulator devices. It is shown that CMMR devices can achieve a 12dB higher IP$_3$ compared to MZI devices. CMMR devices have high second order nonlinearity, while MZI devices' second order nonlinearity is zero. A novel geometry based on dual CMMR modulators is introduced to improve the second order nonlinearity of CMMR modulators.

## 1. Introduction

Optoelectronic signal conversion is the main building block of any analog photonic application. RF signals modulate optical carriers in order to utilize photonic signal processing capabilities. Modulators with good linearity and modulation speeds beyond 100GHz are needed for a variety of analog photonic applications.

Among various materials and technologies, lithium niobate MZI modulators are widely used in analog photonic applications. This is due to their high linearity, as well as the high-speed modulation performance. The electro-optic effect in lithium niobate is extremely fast, and is related to the displacement of electrons with respect to crystal lattices, which has a time constant of less than a femtosecond. There is no material-related bandwidth limitation for an electro-optic modulator made from lithium niobate. However, the modulators that are currently made using this material have limited bandwidth due to the absorption of RF signals or phase mismatch between RF signals and optical signals in the long modulation electrodes of MZI devices [1].

Micro-ring modulators are a class of modulators that use the resonance effect to achieve small form factor devices. Various types of micro-ring modulators have been demonstrated in the past [2-3]. The modulation can be achieved by a change of the resonance frequency of the resonator (FMMR), a change in the absorption or Q of the resonator, or a change in the coupling strength to the micro-ring resonator (CMMR). Fig. 1 shows FMMR and CMMR modulators. FMMR modulators are very compact; however, they have an inherent modulation speed limit that is inversely proportional to resonator Q, since the resonator needs to be "charged" or "discharged" to achieve modulation. Previously, we have experimentally demonstrated FMMR modulators using lithium niobate. [4-5].

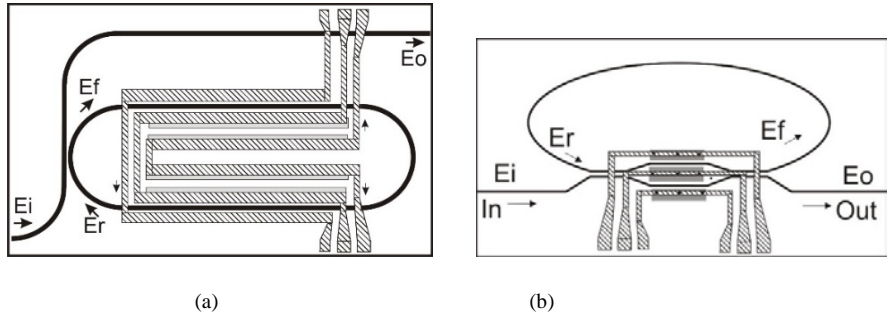

**Fig 1. (a)** FMMR modulator; (b) CMMR modulator

CMMR modulators, on the other hand, are always "charged". Through modulation of the coupling it is possible to achieve a modulator that can completely switch the light on and off at the output while the resonator is always charged [6]. This device is very compact compared to Mach-Zehnder modulators due to high sensitivity caused by the resonator. As opposed to Mach-Zehnder devices, where a complete 180-degree phase shift is needed for switching, in a CMMR modulator, only a few degrees of shift is sufficient to turn the modulator on and off. It has been shown that CMMR modulators have no inherent optical modulation bandwidth limitation imposed by cavity lifetime [7-9]. A limitation will be imposed on the bandwidth by electronics and RF signal losses similar to MZI modulators. However, since the electrodes for CMMR are shorter by a factor of 20-40 compared to MZI devices, it is expected that much higher modulation speeds are feasible using these devices. In this paper we introduce a method to analyze the linearity of CMMR modulators and show in addition to superior bandwidth they also exhibit higher linearity.

Previously, different methods have been introduced to analyze the dynamic performance of micro-ring resonators. Sacher and Poon [7-9] derived the dynamic of modulators for the first time using time domain simulation methods. Their method for the first time showed that CMMR devices do not have a modulation speed limitation, as opposed to FMMR devices. Their calculation method is a time domain method, and it is not easy to calculate the frequency domain response and nonlinearities as needed for many applications. Hong and Enami [12] also calculated the dynamic of micro-ring modulators using the time domain method. Some other small signal analysis frequency domain methods have been introduced more recently [13-14]. Here we provide a frequency domain analysis method that is very simple and is not limited to small signals. The calculation is very fast and requires only a tiny matrix inversion. This calculation method is quite general and can be applied to any resonance-based modulator device. It provides the optical response of the device and can be used to analyze any photonic modulator device including silicon photonic micro-ring modulators.

Using this analysis method, the linearity of these modulators for analog photonic applications are then easily calculated. We will show that the critical $IP_3$ of these modulators is much better compared to MZI devices. We show that CMMR devices have a high second order nonlinearity compared to MZI. This is not critical for many applications since the second harmonic signal can be easily filtered out. We also introduce a dual CMMR device structure that improves the second order nonlinearity.

## 2. Analysis of micro-ring modulators for analog photonics

### 2.1 Analysis of FMMR

In order to model the FMMR modulator, we use the Jacobi–Anger function expansion method of the phase modulated optical signal, similar to what has been done previously to analyze the MZI and similar devices [9].

First we consider a simple FMMR device in which the resonance frequency is shifted when a voltage is applied to the electrodes. Using the notation shown in Fig. 1a. we have:

$$E_i = a_0 e^{-i\omega_o t} . \quad (1)$$

as the input field to the device. Assuming a multitude of frequencies are generated in the micro-ring modulator, we write:

$$E_f = \sum_{n=-\infty}^{\infty} b_n e^{-i(\omega_0 + n\omega_{RF})t}, E_o = \sum_{n=-\infty}^{\infty} d_n e^{-i(\omega_0 + n\omega_{RF})t} . \quad (2)$$

After passing a round trip in the electro-optic micro-ring modulator, the phase modulated generated signals can be expanded using the Jacobi–Anger expansion method:

$$E_r = \sum_{n=-\infty}^{\infty} c_n e^{-i(\omega_0 + n\omega_{RF})t} = \sum_{n=-\infty}^{\infty} \sum_{m=-\infty}^{\infty} J_{m-n}(\beta) b_n \alpha e^{-i((\omega_0 + n\omega_{RF})(t-t_d) + \varphi_{DC})} . \quad (3)$$

where $J$ is the Bessel function, $\beta$ is the modulation depth, $\alpha$ is the round trip loss, and $t_d$ is the round trip delay time.

We introduce the notation:

$$a = \begin{bmatrix} \vdots \\ 0 \\ a_0 \\ 0 \\ \vdots \end{bmatrix}, b = \begin{bmatrix} \vdots \\ b_{-1} \\ b_0 \\ b_1 \\ \vdots \end{bmatrix}, c = \begin{bmatrix} \vdots \\ c_{-1} \\ c_0 \\ c_1 \\ \vdots \end{bmatrix}, d = \begin{bmatrix} \vdots \\ d_{-1} \\ d_0 \\ d_1 \\ \vdots \end{bmatrix} . \quad (4)$$

We can then relate the $c_j$ coefficients to $b_j$ using the matrix identity

$$c = M \cdot b . \quad (5)$$

where M is given by:

$$M = \alpha e^{-i\varphi_{DC}} \begin{bmatrix} \ddots & \vdots & \vdots & \vdots & \cdot^{\cdot} \\ \cdots & J_0(\beta)e^{i2\pi(\omega_o - \omega_{RF})t_d} & J_{-1}(\beta)e^{i2\pi(\omega_o - \omega_{RF})t_d} & J_{-2}(\beta)e^{i2\pi(\omega - \omega_{RF})t_d} & \cdots \\ \cdots & J_1(\beta)e^{i2\pi\omega_o t_d} & J_0(\beta)e^{i2\pi\omega_o t_d} & J_{-1}(\beta)e^{i2\pi\omega_o t_d} & \cdots \\ \cdots & J_2(\beta)e^{i2\pi(\omega_o + \omega_{RF})t_d} & J_1(\beta)e^{i2\pi(\omega_o + \omega_{RF})t_d} & J_0(\beta)e^{i2\pi(\omega_o + \omega_{RF})t_d} & \cdots \\ \cdot^{\cdot} & \vdots & \vdots & \vdots & \ddots \end{bmatrix} . \quad (6)$$

For the directional coupler of the resonator we can write:

$$b = \rho \cdot c + i\tau \cdot a . \quad (7)$$

we can then solve equations 5 and 7 to obtain the coefficient $b_j$ using

$$b = i \cdot \tau \cdot (I - \rho M)^{-1} . a \ . (8)$$

where $\rho$ and $\tau$ are the coupling coefficients of coupler between the micro-ring and the waveguide. Also, the output signal amplitude coefficients are calculated:

$$d = -i \cdot \tau \cdot M \cdot b + \rho . a \ . (9)$$

Fig. 2a and Fig. 2b show the modulated electric field amplitude response of a FMMR for the laser frequency and the first four side band (i.e. coefficient $d_0$, $d_1, d_{-1}, d_2$, $d_{-2}$) for 5 GHz modulation frequency and (a) 50GHz modulation frequency, and (b) as a function of the DC bias phase of the resonator (or the equivalently detuning of the laser frequency from the resonator resonance frequency). The calculation is performed for $\alpha=0.98$ and a coupling factor $r=0.97$ and a ring length of 300 microns. The β value is selected to be 0.0942 by assuming a $V_\pi.L$ of 4V-cm and an applied RF voltage of 2.1 volts to the 300-micron long electrodes of a device.

As can be seen from these figures, the amplitude of the generated sidebands is significantly lower for 50GHz modulation frequency compared to 5GHz, which is an indication of strong frequency dependence for the device. Also, there are peaks in the generated sideband response. The bias phase $\phi_{DC}$ to generate the peak in modulation response varies for the different frequencies. The peaks happen when generated sideband frequency is at resonance. Similar results have been previously obtained using other methods of calculation in the past [6-8].

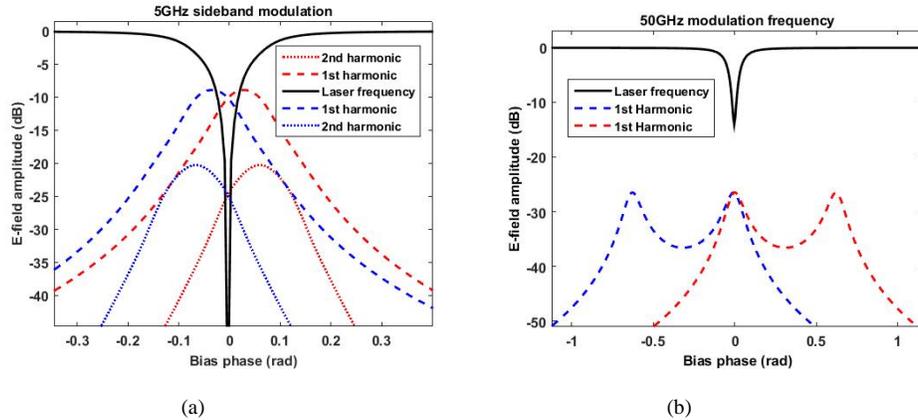

Fig. 2 The generated first harmonic, second harmonic, and fundamental (laser frequency) for (a) 5GHz and (b) 50GHz modulation frequency for the FMMR modulator.

Once the amplitudes of electric-field side-bands are calculated, we can calculate the optical intensity signal frequency response of the modulator by converting the side-band electric field signal amplitude to an intensity signal using the equation

$$I_n = \sum_{m=-\infty}^{\infty} d^*_m . d_{n-m} \ . (10)$$

where $n$ is the $n_{th}$ generated harmonic in the output modulated optical intensity signal at the detector. Notice that here for $n=0$, we get a DC signal.

Figure 3 shows the calculated modulated intensity signal frequency response of a FMMR modulator for the first (*n*=1) and second (*n*=2) harmonics. The response is with respect to input laser power. As can be seen here, the frequency response drops at higher frequency and has a peak at 5GHz. The bias phase was selected such that the laser frequency is 6GHz away from the resonance frequency. As can also be seen in the figure, a large second harmonic signal is generated. It is clear that FMMR devices do not address the high-speed performance needed for RF photonics and we will not further explore these devices in this paper.

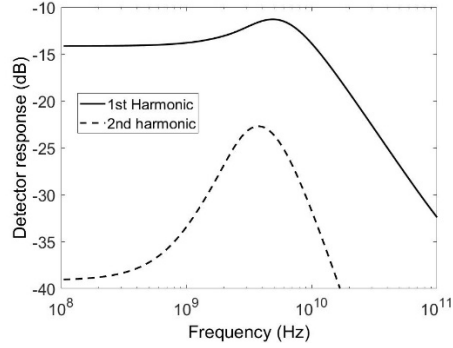

*Fig 3. Calculated detector current frequency response for FMMR modulator for first and second harmonics*

## 2.2 Analysis of CMMR modulators

In order to achieve micro-ring modulators with high-speed performance, one can use CMMR modulators as shown in Fig 1b. It has been shown that in this geometry, there is theoretically no optical related limitation for high-frequency performance [6-8]. Here we develop a simple and quick method to obtain the device frequency response. We also analyze the linearity of these modulators.

Similar to the previous analysis, we can analyze CMMR modulator response using the Jacobi–Anger function expansion method. The MZI section of the micro-ring modulator mixes the optical signals that are received by its two inputs with an RF signal applied to its electrodes. Since there is a feedback loop, the generated side-lobes are fed back into the MZI section and mix again to produce other optical harmonic frequency components. The system can be modeled as follows (using Fig 1b):

$$E_i = a_0 e^{-i\omega_o t} . (11)$$

$$E_f = \sum_{n=-\infty}^{\infty} b_n e^{-i(\omega_0 + n\omega_{RF})t}, E_o = \sum_{n=-\infty}^{\infty} d_n e^{-i(\omega_0 + n\omega_{RF})t} . (12)$$

$$E_r = \sum_{n=-\infty}^{\infty} c_n e^{-i(\omega_0 + n\omega_{RF})t} = \sum_{n=-\infty}^{\infty} \alpha b_n e^{-i(\omega_0 + n\omega_{RF})(t-t_d)} . (13)$$

Similar to the FMMR analysis, $\alpha$ is the round trip loss in the micro-ring resonator. Again, using the notation

$$a = \begin{bmatrix} \vdots \\ 0 \\ a_0 \\ 0 \\ \vdots \end{bmatrix}, b = \begin{bmatrix} \vdots \\ b_{-1} \\ b_0 \\ b_1 \\ \vdots \end{bmatrix}, c = \begin{bmatrix} \vdots \\ c_{-1} \\ c_0 \\ c_1 \\ \vdots \end{bmatrix}, d = \begin{bmatrix} \vdots \\ d_{-1} \\ d_0 \\ d_1 \\ \vdots \end{bmatrix}. \quad (14)$$

we can write

$$c = M_3 \cdot b. \quad (15)$$

where here $M_3$ is a diagonal matrix. Its element are constant in time and can be easily written using (13). Assuming a multi-tone input optical signal is fed to the MZI section from the feedback section of the device, each arm of MZI section produces side-bands related to its modulation RF frequency that are related to its input optical signal via the Bessel functions. In MZI section there are two -3dB couplers, two modulating paths for the optical signal, as well as two inputs and two outputs. We should calculate the contribution of each of the two input arms to each of the two output arms via the two paths. The contribution from $E_r$ to $E_f$ after passing two modulating paths can be written as:

$$E_f^r = \sum_{n=-\infty}^{\infty} b_n^r e^{-i(\omega_0 + n\omega_{RF})t} = \frac{e^{i\frac{\varphi_{DC}}{2}}}{2} \sum_{n=-\infty}^{\infty} \sum_{m=-\infty}^{\infty} (i)^{m-n} J_{m-n}(\beta) c_n e^{-i(\omega_0 + n\omega_{RF})t}$$
$$+ \frac{i \cdot e^{-i\frac{\varphi_{DC}}{2}}}{2} \sum_{n=-\infty}^{\infty} \sum_{m=-\infty}^{\infty} (i)^{m-n} J_{m-n}(-\beta) c_n e^{-i(\omega_0 + n\omega_{RF})t} \quad (16)$$

$\beta$ is the modulation depth of the phase modulation sections of MZI and is given by $\beta = (V/V\pi)$, $\varphi_{DC}$ is the DC bias phase difference on the two arms of the MZI section, and $J$ is the Bessel function. We can write this equation using matrix notation:

$$b^r = M_1 \cdot c. \quad (17)$$

Similarly, the contribution from $E_r$ to $E_o$ after passing two paths of the Mach-Zehnder can be written as:

$$E_o^r = \sum_{n=-\infty}^{\infty} d_n^r e^{-i(\omega_0 + n\omega_{RF})t} = \frac{-ie^{i\varphi_{DC}}}{2} \sum_{n=-\infty}^{\infty} \sum_{m=-\infty}^{\infty} (i)^{m-n} J_{m-n}(\beta) c_n e^{-i(\omega_0 + n\omega_{RF})t}$$
$$+ \frac{e^{-i\varphi_{DC}}}{2} \sum_{n=-\infty}^{\infty} \sum_{m=-\infty}^{\infty} (i)^{m-n} J_{m-n}(-\beta) c_n e^{-i(\omega_0 + n\omega_{RF})t} \quad (18)$$

We can write this equation using matrix notation:

$$d^R = M_2 \cdot c. \quad (19)$$

Next, we need to add the contribution from input signal $E_i$ to $E_F$ and $E_R$. The transfer function from $E_i$ to $E_F$ and $E_R$ is similar to equations 15 and 17, and is not repeated here. The results can be summarized by:

$$b = b^r + b^i = M_1 \cdot c + M_2 \cdot a \quad (20)$$

$$d = d^i + d^r = M_2 \cdot a + M_1 \cdot c \quad (21)$$

These matrices relate the signal coefficient of $E_F$ and $E_O$ to $E_R$ and $E_i$. They are similar to terms $\tau$ and $\rho$ for a simple directional coupler, but have Bessel function elements due to an electro-optic modulation section. By solving equations 15 and 20, one obtains $b_j$ values:

$$b = (I - M_1.M_3)^{-1} M_2 \cdot a \quad (22)$$

The output signal coefficients is then obtained by using equation 21:

$$d = M_2 a + M_1.M_3 \cdot b \quad (23)$$

Here we use this analysis method to investigate a few different conditions for the system response.

Fig. 4 shows the laser frequency, first and second harmonics (i.e., coefficient $d_0, d_1, d_{-1}, d_2, d_{-2}$), amplitude coefficients generated for electric field side-bands in a CMMR modulator for 5GHz and 50GHz as a function of DC bias point $\varphi_{DC}$. As opposed to the FMMR devices discussed above, in this case, between zero coupling and critical coupling bias points, there are slight changes in the amplitude of the first harmonic generated electric Field signals for 5GHz and 50 GHz modulation frequencies. In this case, for certain bias values, the generated side-bands at 50GHz are even slightly higher than the generated signal at 5GHz. There are also second harmonic generated signals and higher harmonics as shown in Figure 4. These results theoretically demonstrate that CMMR modulators do not have high-frequency roll-off, as opposed to FMMR modulators.

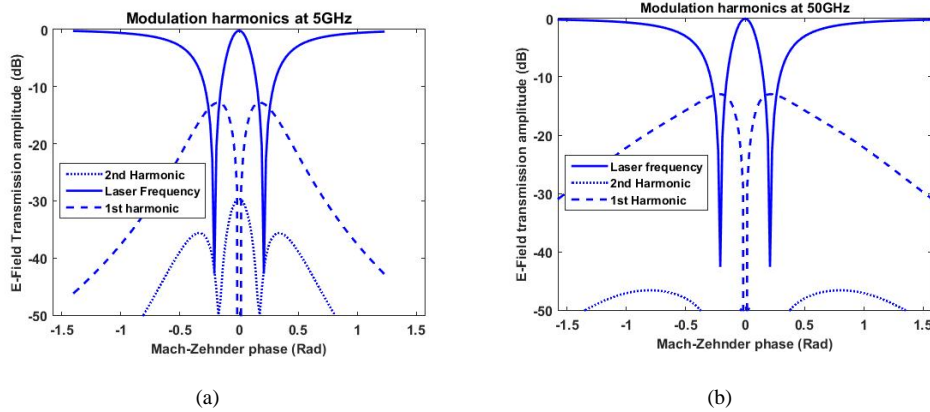

Fig. 4 (a) The generated first harmonic, second harmonic, and fundamental (laser frequency) for (a) 5GHz and (b) 50GHz modulation frequency for CMMR.

Similar to the previous case, we calculate the intensity signal that is generated after detection for CMMR devices using equation 10. Fig. 5 shows the intensity signal frequency response of the modulator for two different bias values. The signal is relative to input laser power, As can be seen here, depending on the selected bias value it is possible to obtain different frequency responses for the system. As can be seen later, we will use a bias point between zero coupling and critical coupling for linear modulators. Hence, the low frequency response for linear modulators devices will be similar to the case where the bias is 0.15 in Fig. 5. Similar results have been published previously for the frequency response of CMMR modulators. [6-8]

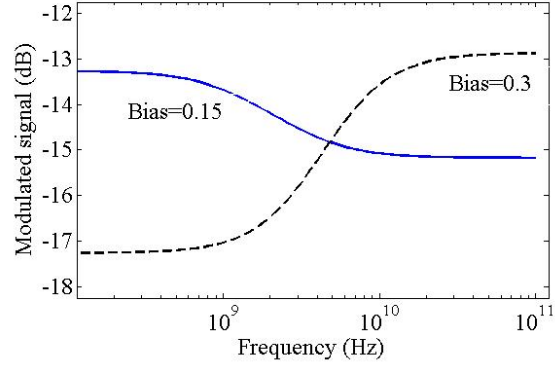

Fig. 5. The frequency response of CMMR for two different bias values

## 3. Linearity of CMMR modulators

For analog photonic applications, the linearity of a modulator is very critical. Here we analyze the linearity of these modulators. Fig. 6 shows the calculated fundamental and the second and third harmonic frequency response of a CMMR modulator as function of modulation frequency. These signal levels are with respect to input laser power. The bias phase is appropriately selected between zero coupling and critical coupling point and is equal to 0.12 radian. Similar to MZI modulators, this bias phase results in high first harmonic power and low higher harmonic signal power. The second harmonic is the dominant distortion harmonic for a CMMR device. The third harmonic is very low for a CMMR device. As a comparison, for a standard MZI device, there is no second harmonic signal at the quadrature bias point and the third harmonic signal is 35 dB lower than first harmonic signal for similar fundamental signal power levels (i.e.~ -50 dB in Fig. 6). Hence the CMMR has lower third harmonic distortion compared to MZI for high-frequency modulation speeds. For many applications, the second harmonic distortion is not important since it can be filtered out. Hence, CMMR devices may not only allow significantly wideband operations, but can also provide better linearity.

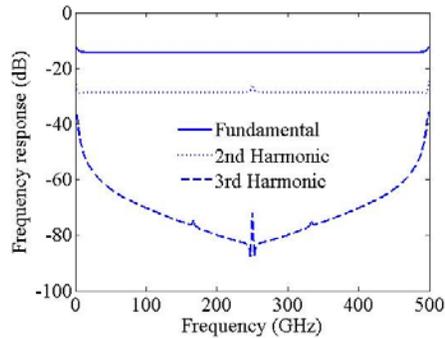

Fig. 6 The first, second, and third harmonic for a CMMR for different modulation frequencies

For specific wideband applications where the second harmonic might be critical, we propose the device structure that is shown in Fig. 7(a). In this device, two CMMR modulators are used. By careful selection of the RF signal that is applied to the electrodes of this modulator, it is possible to eliminate the second harmonic side-band generated in CMMR modulators. In order to achieve this, the modulated RF signal applied to the second micro-ring device electrodes must be phase shifted by 90 degrees with respect to the first device. Also, the bias point for the combination of modulated signals before the output coupler should be 90 degrees out of phase with respect to each other.

Figure 7(b) shows the calculated first, second, and third harmonic signal levels with respect to input optical power after detection for a dual CMMR (DCMMR) device. In this device the second harmonic is completely eliminated in the modulated electric field signal. However, there is still a second harmonic signal in the intensity signal. However, the amplitude of the second harmonic intensity signal is significantly lower for DCMMR modulators compared to single-CMMR modulators. One disadvantage of DCMMR devices is inherent optical loss of the device. An additional inherent -9dB loss exists in DCMMR devices.

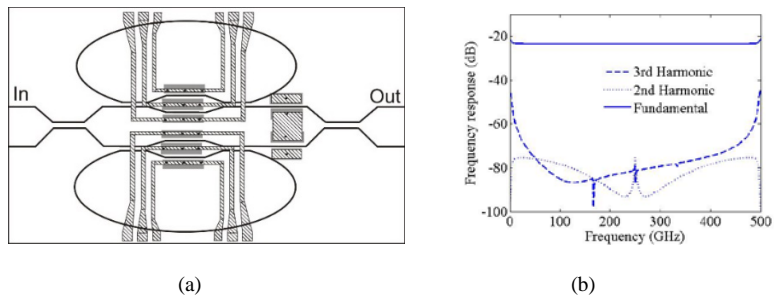

(a)  (b)

Fig. 7. Proposed device to achieve high linearity based on DCMMR modulators; (b) Calculated fundamental, second harmonic, and third harmonic signal distortion levels for DCMMR

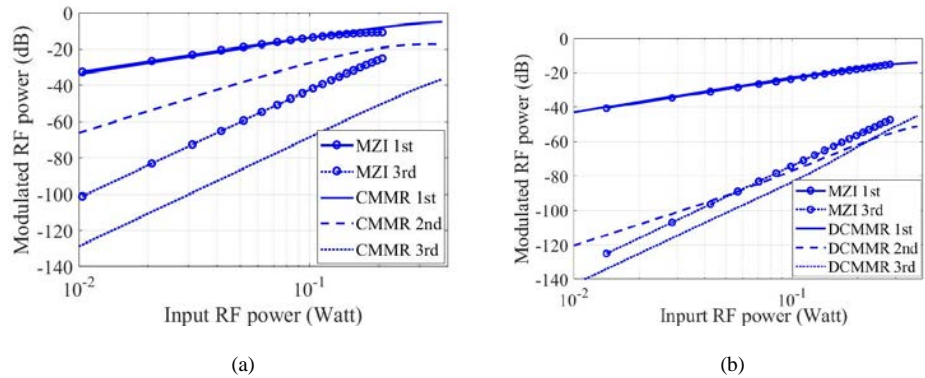

(a)  (b)

Fig. 8. The SFDR and its comparison with a simple Mach-Zehnder modulator for double CMMR (solid line) and Mach-Zehnder (dashed line) modulators

Figure 8 shows the fundamental, second harmonic, and third harmonic levels for CMMR and DCMMR devices and compares the results with MZI devices. The results are plotted for a modulation frequency of 100GHz. For single-CMMR devices, the third order harmonic power is approximately 25 dB lower than MZI devices for similar fundamental signal power levels. This translates to a 12.5 dB improvement in the $IP_3$ compared to MZI devices. For lower frequencies the improvement is less. For example, at 50GHz the improvement is 10dB in $IP_3$ and at 200GHz the improvement is 17dB for $IP_3$. For very low frequencies (less than 20GHz) the third harmonic distortion $IP_3$ is similar to MZI.

For dual-CMMR, due to additional inherent insertion loss devices, the $IP_3$ and third harmonic distortion is similar, slightly worse, or slightly better depending on the frequency compared to MZI devices.

One issue with these devices is the low frequency distortions that are caused by the resonator cavity dynamics, which also repeats once the modulation frequency reaches the free spectral range of the device (i.e., 0Hz and 500GHz in our device). This problem can be easily solved using the methods previously described initially by Popovic [15], and was re-iterated in a more recent publication by Kodanev and Orenstein[10]. Basically, another port is added to compensate for energy loss from the micro-resonator which completely eliminate the cavity dynamics in the modulation response. We expect a frequency independent $IP_3$ improvement of 12 dB to be practical using these designs.

4.  **Conclusion**

We have developed a novel simple method to analyze the micro-ring modulator frequency response. We applied this method to resonance FMMR modulators as well as CMMR modulators. The frequency response results obtained are similar to the results obtained previously using the time domain simulation method [7-9]. The analysis shows that CMMRs do not have a modulation speed limit imposed by photon lifetime in the resonator.

This theory was then applied to analyze the linearity of micro-ring modulators. Single-CMMR devices and double-CMMR devices were analyzed for their linearity and were compared to MZI devices. The results show that CMMRs have superior performance due to much lower third order distortion compared to MZI devices. However, CMMR devices have a large second order distortion that might be problematic for some applications. Double-CMMR devices have low second order distortion, but the improvement in third order distortion is compensated by additional inherent insertion loss of the device. There is low frequency distortion in the device caused by the cavity dynamics that can be fixed by using additional ports in the device to as was shown in [11] and [15].

**Acknowledgement:** This work is supported by the NASA STTR Program under contract number T8.02-9806 (STTR 2016-1).